\documentclass[10pt,%
aip,
 amsmath,amssymb,
reprint,%
author-numerical,
floatfix,
]{revtex4-1}
\usepackage{graphicx}% Include figure files
\usepackage{dcolumn}% Align table columns on decimal point
\usepackage{bm}% bold math
\usepackage{xcolor}
\usepackage{epstopdf}
\usepackage{url}
\begin{document}

%\preprint{AIP/123-QED}

\title{Microscopic nature of the asymmetric hysteresis in the insulator-metal transition of VO$_2$ revealed by spectroscopic ellipsometry. }

\author{Ievgen Voloshenko}
\affiliation{
\mbox{1.~Physikalisches Institut, Universit\"at Stuttgart, Pfaffenwaldring 57, 70550 Stuttgart, Germany}
}%
\author{Florian Kuhl}
\affiliation{
Institute for Experimental Physics I and Center for Materials Research (LaMa), Justus Liebig Universit\"at Giessen, Heinrich-Buff-Ring 16, 35392 Giessen, Germany
}%
\author{Bruno Gompf}%
\affiliation{
\mbox{1.~Physikalisches Institut, Universit\"at Stuttgart, Pfaffenwaldring 57, 70550 Stuttgart, Germany}
}%
%\email{b.gompf@pi1.physik.uni-stuttgart.de}

\author{Angelika Polity}
\affiliation{
Institute for Experimental Physics I and Center for Materials Research (LaMa), Justus Liebig Universit\"at Giessen, Heinrich-Buff-Ring 16, 35392 Giessen, Germany
}%
\author{Gabriel Schnoering}%
\affiliation{
\mbox{1.~Physikalisches Institut, Universit\"at Stuttgart, Pfaffenwaldring 57, 70550 Stuttgart, Germany}
}%
\author{Audrey Berrier}%
\affiliation{
\mbox{1.~Physikalisches Institut, Universit\"at Stuttgart, Pfaffenwaldring 57, 70550 Stuttgart, Germany}
}%
\author{Martin Dressel}%
\affiliation{
\mbox{1.~Physikalisches Institut, Universit\"at Stuttgart, Pfaffenwaldring 57, 70550 Stuttgart, Germany}
}%

\newcommand{\cmmnt}[1]{\ignorespaces}

\date{\today}% It is always \today, today,
\begin{abstract}
Systematic spectroscopic ellipsometry investigations have been performed in order to elucidate the asymmetric  insulator-to-metal transition in thin VO$_2$ films. The comprehensive analysis of the obtained macroscopic optical response yields a hysteretic behavior, and in particular its asymmetry, when performed in the framework of an anisotropic effective medium approximation taking into account the volume fraction of the metal inclusions as well as their shape. We reveal microscopic details of the percolation transition, namely that the shape of the metal inclusions goes through several plateaus, as seen in the evolution of the shape factor on both sides of the transition region and resulting in different critical volume fractions at the transition for the heating and cooling cycles.

% Valid PACS numbers may be entered using the \verb+\pacs{#1}+ command.
\end{abstract}


\maketitle

Vanadium dioxide undergoes an insulator-to-metal transition (IMT) at around $68^\circ$C that is accompanied by a structural transformation from a dimerized, monoclinic (P2$_1$/c) lattice below the critical temperature to a (P4$_2$/mnm) structure above.\cite{Morin} The coupling of electronic and structural degrees of freedom has engendered the ongoing discussion about the origin of the transition, whether it is a structural Peierls transition, an electronic Mott-Hubbard transition or a combination of both.

Additional complexity is introduced by the hysteretic behavior of the transition, observed for example in resistivity\cite{Rozen,Sharoni} and Hall measurements,\cite{Yamin,Hall} optical spectroscopy in different frequency ranges,\cite{Barker, Hood, Hui} and structural investigations.\cite{Yao,Koethe}  Advanced microscopic techniques, such as nanoscale x-ray microscopy\cite{Kumar, Qazilbash2} and near-field optical microscopy,\cite{Qazilbash2} reveal the percolating nature of the transition with growing metallic domains in the insulating matrix as temperatures rises. The width of hysteresis depends on the preparation procedure,\cite{method} the substrate and its orientation.\cite{substrate,Li,Zhu} In particular, the width of the transition is linked to the quality of the material, with thinner and steeper transitions for better materials. Moreover, an asymmetric hysteresis is observed both in single crystals and thin films, with a steeper transition from the metallic to the insulating phase than {\it  vice versa}.\cite{Tobi,Ji,Xiong} Although {\it ab-initio} calculations reproduce the main characteristics of the IMT in VO${_2}$, \cite{Gatti,Eyert, Eyert2} they neither include the hysteretic nature nor its asymmetry, which is governed by microscopic and thermodynamic processes.\cite{PhysRevB.83.235102}

In principle, percolation transitions can be described by a Bruggeman effective medium approximation (BEMA), where the effective properties of the composite are obtained from the properties of its constituents. However, conventional isotropic BEMA models face certain well known limitations. The critical volume fraction is coupled to the shape parameter of the inclusions, which in the case of spherical inclusions leads to a critical volume fraction where the IMT occurs inherently at 33$\%$. For VO$_2$, it was shown that the simple isotropic BEMA model fails to reproduce optical and transport properties in the vicinity of the transition;\cite{Choi, James} calling for an improvement of the BEMA model.

\begin{figure}[b]
  \centering
  \includegraphics[width=0.8\columnwidth]{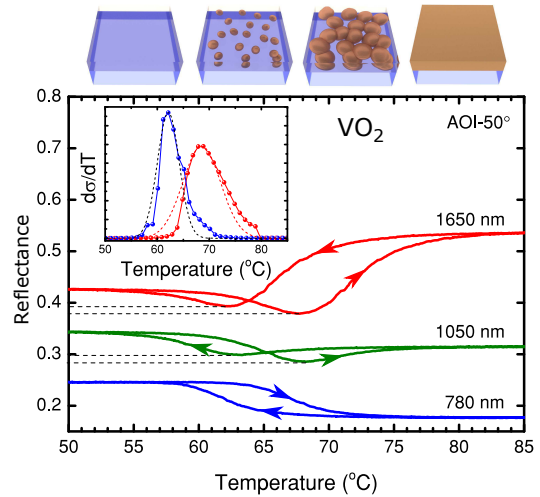}
  \caption{Temperature dependent reflectance of VO$_2$ for different wavelengths $\lambda$. The dashed lines indicate the minimal values. The inset shows the derivative of the electrical conductance with respect to temperature for heating (red) and cooling (blue) cycles. On top, the development of percolating domains is illustrated.}
  \label{fig:fig1}
\end{figure}

In this Letter, we demonstrate that the IMT of VO$_2$ can be described in all details by an anisotropic BEMA with varying both the filling and the shape factor. The shape factor departs from spherical inclusions around the transition region and has different values for heating and cooling. The critical volume fraction is therefore different leading to the pronounced asymmetric hysteretic behavior. In the framework of the BEMA, the macroscopic optical response is consistently reproduced over the whole transition region and microscopic details of the percolation transition are unveiled.

We investigated vanadium dioxide thin films deposited by radio frequency sputtering using a metallic V target in a reactive process with an Ar/O$_{2}$ gas mixture on an Al$_2$O$_3$ substrate of 510 $\mu$m thickness, with the $c$-axis (0001) perpendicular to the plane. During deposition, the gas pressure was about 3.4$\times$10$^{-3}$ mbar and the ratio of the mass flows O$_2$/Ar was adjusted to 3.1$\%$. The growth temperature was kept at 550 $^\circ$C over the full deposition period of 60~min.\cite{Dietrich} By ellipsometric measurements we determined the thickness to 105$\pm$3 nm, with a roughness of about 15$\pm$0.25 nm, in good agreement with the root-mean-square roughness obtained by AFM.

For spectroscopic measurements we utilized a Woollam RC2-UI ellipsometer since its wide spectral range ($\lambda=210$ to 1690~nm) allows determining the dielectric properties of the sample with high precision by modeling measured ellipsometric parameters $\Psi$ and $\Delta$.\cite{Fujiwara} A heating stage enables temperature-dependent measurements with 0.1$^\circ$C resolution; the temperature is measured in-situ by a sensor.
During the experiments heating and cooling rates were kept at 3$^\circ$C/min with an acquisition time of 1~s. The reported temperature refers to the VO$_2$ film estimated by taking into account the heat conductivity and convective heat transfer coefficients of the respective materials. It was checked that the temperature difference between stage and film stays smaller than 1$^{\circ}$C over the whole temperature range.
Measuring the electrical impedance at 11~MHz allows us to determine the IMT phase transition  upon cooling and heating. As plotted in the inset of Fig.~\ref{fig:fig1}, we observe a significant asymmetry, having sharper and more abrupt metal-to-insulator transition while cooling.
The critical temperatures for these VO$_2$ films are $T_{\rm heat}= 68^\circ$C and $T_{\rm cool}= 61.5^\circ$C for heating and
cooling cycles, respectively.

Figure \ref{fig:fig1} presents the reflectance of the VO$_2$ films at $\lambda=1650$, 1050 and 780~nm, measured at an angle of incidence of 50$^\circ$ for heating-cooling cycles.
We observe a complex temperature dynamics  depending on temperature and wavelength range. For smaller $\lambda$ the reflectance decreases with temperature. Above 900~nm free-charge-carrier contributions become prominent as metallic clusters form. Hence the reflectance first decreases, goes through a minimum around the critical temperature and then rises again. Such a temperature behavior reveals the presence of a percolation transition. The reflectance decreases due to absorption\cite{Hoevel} that arises from the conductive clusters generating localized plasmon modes; this also explains the spectral modifications of the reflectance curves. The minima in reflectance (dashed lines in Fig.~\ref{fig:fig1}) occur at different values upon heating or cooling, respectively, providing evidence for distinct optical properties (e.g.\ spectral position of the localized plasmonic resonance) at the transition temperature.

\begin{figure}[t]
  \centering
  \includegraphics[width=0.8\columnwidth]{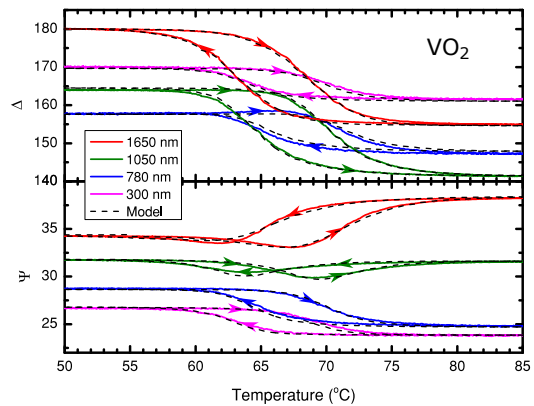}
  \caption{Ellipsometric parameters $\Psi$ and $\Delta$ measured during cooling and heating at different wavelengths. The black dashed curve represents the fit by the BEMA model. Notice the asymmetric behavior of $\Psi$ for $\lambda=1650$ and 1050 nm.}
  \label{fig:fig2}
\end{figure}

To further investigate the asymmetry of the transition, we consider the raw data of the ellipsometric measurement, i.e.\ the parameters $\Psi$ and $\Delta$ that define the complex reflectance coefficient $\rho=r_p/r_s=\tan\{\Psi\}\exp\{i\Delta\}$ and allow us to extract the complex dielectric properties. A proper model accounts for the presence of the substrate and  simultaneously fits both ellipsometric parameters as a function of the wavelength and temperature.

The Bruggeman effective medium approximation describes the material from the perspective of a composite, with inclusions of given shape (clusters of metallic phase) embedded in an isotropic matrix with known dielectric properties (insulating phase):\cite{choy}
\begin{align}
  f\frac{\varepsilon_i - \varepsilon_{\rm eff}}{\varepsilon_{\rm eff}+L(\varepsilon_{i} - \varepsilon_{\rm eff})}+ (1-f)\frac{\varepsilon_{m} -\varepsilon_{\rm eff}} {\varepsilon_{\rm eff} +L(\varepsilon_{m} -\varepsilon_{\rm eff})}=0,
  \label{eq:eq1}
\end{align}
where $\varepsilon_i$, $\varepsilon_m$ are the complex dielectric functions of the insulating and metallic phases, $f$ is the volume fraction of the metallic inclusions and $L$ is their shape factor. $\varepsilon_{\rm eff}$ is the effective dielectric function of the composite.
A Cauchy model accounts for the optical properties of the Al$_2$O$_3$ substrate for both polarizations. Since it is double-side-polished, its birefringence induces strong depolarization from incoherent back reflection at oblique incidence,\cite{Backrefl} which is taken into account numerically in order to obtain the exact properties of the film.

The BEMA model is based on the assumption that the optical behavior of the composite is fully determined by its constituents, i.e. the insulating and metallic phases. In a first step we thus determined their optical properties at $T=23$ and $90^\circ$C, since at these temperatures the films form a homogeneous single phase, respectively.
We start with the insulating phase. Despite the polycrystalline nature of the film, the strain induced by the substrate results in uniaxial properties of the film. To obtain the dielectric response of both in- and out-of-plane optical axes, we performed  ellipsometric measurements by varying the angle of incidence from 25$^\circ$ to 65$^\circ$. The in-plane response is modelled with a Tauc-Lorentz function with a band gap of $E_g=0.56$~eV and two Gaussian functions at energies $E_1=3.3$~eV and $E_2=8.3$~eV to account for the interband transitions;
this is in accord with previous reports.\cite{Tomczak,Huffman}
The out-of-plane direction uses the same model with a band gap of $E_g=0.58$ eV and Gaussians of energies at $E_3=2.6$~eV and $E_4=8.5$~eV.

The properties of the metallic phase are reproduced by adding a Drude term to the previous model to account for the free charge carriers contribution. The band gap closes and the Tauc-Lorentz terms become simple Lorentzian oscillators along both axes with vanishingly small amplitude. For the charge-carriers we determine a density of $9.7 \times 10^{21}\ \mathrm{cm^{-1}}$, their mobility is $0.5\ \mathrm{cm^2\ V^{-1}\ s^{-1}}$ and their effective mass ${m^*}/{m_e}= 1.4$;
other experiments \cite{BasovIR,Hall} confirm our findings.
The interband transition shifts with respect to the insulating phase, both for in-plane and out-of-plane axes with ($E_1=3.27$ eV, $E_2=8.34$ eV) and ($E_3=3.11$ eV, $E_4=8.12$ eV), respectively. While in-plane, the interband transition is slightly reduced, for the out-of-plane direction we observe a strong increase governed by induced strain over the transition. The normal-incidence spectra simulated with these two models perfectly describes our data obtained  for both phases by Fourier-transform infrared spectroscopy.\cite{Tobi}

With the properties of the insulating and metallic phases fixed,
there remain 3 free fit parameters in the temperature dependent BEMA model; the  most important ones are the volume fraction $f$ and the shape factor $L$ of the metallic inclusions.
The surface roughness is a third parameter. The roughness of the film is approximated by an effective medium, i.e.\ a layer that combines the optical properties of the film together with air as a 50$\%$ mixture.
We determine a roughness of 15$\pm$0.25 nm which increases by approximately 5 nm when passing through the transition.
It has been suggested \cite{Anisotropic} that the roughness varies due to the lattice expansion at the transition.

The shape factor $L$ is a correction of the local field accounting for the geometry of the particle. Inclusions smaller than the wavelength can be approximated by ellipsoids with fixed revolution axis having a shape factor $L=(L_{\perp},L_{\parallel},L_{\parallel})$. The out-of-plane term of the shape factor can be expressed as:\cite{Hulst}
\begin{align}
  L_{\perp}=\frac{1+r^2}{r^2}\left(1-\frac{1}{r}\tan^{-1}(r)\right) \quad ,
  \label{eq:eq2}
\end{align}
where $r^2=x^2/z^2-1$, the ratio of the in-plane and out-of-plane radii of the ellipsoid, $x$ and $z$, respectively. The in-plane part is given by $L_{\parallel}=\frac{1}{2}(1-L_{\perp})$.
For a sphere one finds $L=(\frac{1}{3},\frac{1}{3},\frac{1}{3})$,
for a needle $(0,\frac{1}{2},\frac{1}{2})$, and for a disc $(1,0,0)$.\cite{Taylor}
In the BEMA model, the shape of the inclusions defines the critical volume fraction, at which a material percolates; in the simplest case of spheres it is $33\%$ .\cite{choy}
The model assumes that the metallic cluster -- although growing in size -- are always much smaller than the probing wavelength. This assumption is validated by near-field measurements,\cite{Qazilbash2} which demonstrate that these clusters reach more than hundreds of nanometers in cross-section; nevertheless they remain well separated. The shape of the clusters, however, becomes crucial and has to be taken into account;
they are  well approximated by ellipsoids. In other words, since the BEMA assumptions are not violated we can now adjust the shape factor to optimize the description of our observations.

\begin{figure}[t]
  \centering
  \includegraphics[width=0.8\columnwidth]{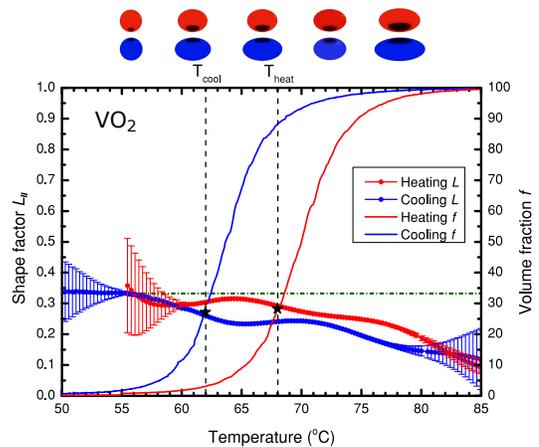}
  \caption{Temperature variation of the shape factor together with associated volume fractions for heating and cooling cycles. The error bars indicate the experimental uncertainty ($1\sigma$). Spherical inclusions are visualized by the green dashed line and stars represent the critical volume fraction at the transition temperature predicted by percolation theory.}
  \label{fig:fig3}
\end{figure}

The resulting fit of $\Psi$ and $\Delta$ with such a model is displayed in Fig.~\ref{fig:fig2} for several wavelengths; note that we keep the fitting procedure consistent and made only minimal variations. Our model reproduces well the measurements over a wide spectral range from ultraviolet to near-infrared; the deviations are strongest around the critical temperatures. In comparison to a static model with spherical inclusions, the maximal mean-square error at the critical temperature is four times smaller and the temperature range with high values is dramatically reduced.
The remaining deviations are attributed to the so-called Lifshitz tails, arising from phase fluctuations,\cite{Brouers} which the BEMA model does not account for.

Fig.~\ref{fig:fig3} displays the in-plane shape factor $L_{\parallel}$ together with the volume fraction $f$ of the metallic clusters. The shape factor represents the mean value of seven different measurements. The experimental uncertainties increase when the insulating or the metallic phase is reached because a precise determination of the shape factor is not possible in the single-phase limit.
It is important to notice that $L$ exhibits a hysteresis. Via Eq.~\ref{eq:eq2} this implies that also the shape of the clusters is different for the heating and cooling cycles, as sketched by the red and blue spheroids on top of Fig.~\ref{fig:fig3}.
Slightly above room temperature, where the filling fraction is still negligible, spherical metallic clusters start to form. 
Approaching the critical temperature, the shape factor decreases until it reaches the first plateau at $L_\parallel =0.25$. 
Here the clusters spread in plane approaching a disc-like shape; 
in agreement with previous observations.\cite{Kim} 
Above $T=72^\circ$C, $L_\parallel$ drops to 0.14. 
The shape factor does not vary linearly with temperature:
a valley above the critical temperature is followed by a broad peak for both cycles.
It is worth noticing that the evolving shape of the clusters with temperature may explain the departure in the conductivity derivatives from the expected gaussian shape as seen in the inset of Fig. \ref{fig:fig1}.
% Slightly above $70^\circ$C the plateau acquires almost the same value during heating and cooling.  Most important, however,
% the changes in $L_{\parallel}$ occur at different temperatures.
% This is the reason for the asymmetric hysteresis behavior.

We now can determine the critical shape and volume fraction at the
transition temperatures obtained by impedance measurements.
The geometry and filling fractions of the metallic inclusions are different at their respective critical temperatures indicated by dashed lines in Fig.~\ref{fig:fig3}: during heating $f_c^h=28.8\%$ and $L_\parallel^h=0.285$, resulting in $z/x=5/7$;
and $f_c^c = 26.2\%$ and $L_\parallel^c=0.265$ during cooling, 
corresponding to an axial ratio of $2/3$. 
This difference in shape and volume fraction leads to 
different effective dielectric constants, which adequately describes the asymmetry of the hysteteric behavior observed in the optical response.

It is interesting to note that percolation theory predicts\cite{Snyder} that aggregated ellipsoids of such shape percolate at critical volume fractions of $28.2\%$ and $27.1\%$, respectively (stars on the Fig.~\ref{fig:fig3}); in good accord with our results.
The excellent agreement between anisotropic BEMA and percolation theory
at the critical fraction strongly supports our procedure. 
We want to emphasize that the model solely considers a geometric percolation of purely metallic and completely insulating states; there are no indications of an intermediate electronic states in VO$_2$.

In summary, a detailed analysis of our temperature-dependent ellipsometric measurements of VO$_2$ films in the framework of an anisotropic Bruggeman effective medium approximation allows us to determine the microscopic geometry of the constituents and to explain the asymmetric percolation transition in VO$_2$. In the course of the IMT the metallic inclusions continuously change in shape flattening at large temperatures and acquire distinct shapes at critical temperatures during heating and cooling. This explains the observed asymmetry of the hysteresis loop. We are now able to describe accurately the temperature dependent dielectric constant when passing through the phase transition.

Financial support is provided by the DFG via the GRK (Research Training Group) 2204 `Substitute Materials for sustainable Energy Technologies'.

\bibliography{library}

\end{document}